\pdfoutput=1 
\documentclass[10pt,conference]{IEEEtran}
\IEEEoverridecommandlockouts
\usepackage{cite}
\usepackage{amsmath,amssymb,amsfonts}
\usepackage{algorithmic}
\usepackage{graphicx}
\usepackage{textcomp}
\usepackage{xcolor}
\def\BibTeX{{\rm B\kern-.05em{\sc i\kern-.025em b}\kern-.08em
    T\kern-.1667em\lower.7ex\hbox{E}\kern-.125emX}}
\begin{document}

\title{Embedding Learning in Hybrid Quantum-Classical Neural Networks}

\author{\IEEEauthorblockN{Minzhao Liu\IEEEauthorrefmark{1}\IEEEauthorrefmark{2}\IEEEauthorrefmark{9}, Junyu Liu\IEEEauthorrefmark{3}\IEEEauthorrefmark{4}\IEEEauthorrefmark{5}\IEEEauthorrefmark{10}, Rui Liu\IEEEauthorrefmark{6}, Henry Makhanov\IEEEauthorrefmark{7}, Danylo Lykov\IEEEauthorrefmark{2}\IEEEauthorrefmark{6}, Anuj Apte\IEEEauthorrefmark{1}\IEEEauthorrefmark{5}, Yuri Alexeev\IEEEauthorrefmark{2}\IEEEauthorrefmark{6}}
\IEEEauthorblockA{\IEEEauthorrefmark{1}Department of Physics, The University of Chicago, Chicago, IL 60637, USA}
\IEEEauthorblockA{\IEEEauthorrefmark{2}Computational Science Division, Argonne National Laboratory, Lemont, IL 60439, USA}
\IEEEauthorblockA{\IEEEauthorrefmark{5}Kadanoff Center for Theoretical Physics, The University of Chicago, Chicago, IL 60637, USA}
\IEEEauthorblockA{\IEEEauthorrefmark{3}Pritzker School of Molecular Engineering, The University of Chicago, Chicago, IL 60637, USA}
\IEEEauthorblockA{\IEEEauthorrefmark{4}Chicago Quantum Exchange, Chicago, IL 60637, USA}
\IEEEauthorblockA{\IEEEauthorrefmark{6}Department of Computer Science, The University of Chicago, Chicago, IL 60637, USA}
\IEEEauthorblockA{\IEEEauthorrefmark{7}Department of Computer Science, The University of Texas at Austin, Austin, TX 78712, USA}
\IEEEauthorblockA{\IEEEauthorrefmark{10}qBraid Co., Harper Court 5235, Chicago, IL 60615, USA}
E-mail: \IEEEauthorrefmark{9}mliu6@uchicago.edu}

\maketitle

\begin{abstract}
Quantum embedding learning is an important step in the application of quantum machine learning to classical data. In this paper we propose a quantum few-shot embedding learning paradigm, which learns embeddings useful for training downstream quantum machine learning tasks. Crucially, we identify the \textit{circuit bypass problem} in hybrid neural networks, where learned classical parameters do not utilize the Hilbert space efficiently. We observe that the few-shot learned embeddings generalize to unseen classes and suffer less from the circuit bypass problem compared with other approaches.
\end{abstract}


\section{Introduction}

Quantum computing has been investigated in the context of machine learning in recent years and is argued to hold great potential in augmenting existing machine learning algorithms \cite{Du, Abbas, Huang, Cong}. Hardcore proofs of quantum advantage are demonstrated in approaches that utilize quantum inputs, including quantum convolutional neural networks for quantum error correction and quantum phase recognition \cite{Cong} and a quantum kernel method that accelerates linear algebra subroutines \cite{qsvm, hhl}. On the other hand, for classical inputs, parameterized quantum circuits (PQCs) are often utilized for encoding classical information, followed by a variational processing circuit and a measurement circuit for classification \cite{schuld}. It is believed that such models might be more expressive and trainable than their classical counterparts \cite{Abbas}. Numerous studies have examined this encode-process-measure paradigm and argued for potential quantum advantage \cite{Havlicek}.

Recently, the search for hybrid quantum-classical machine learning architectures that can boost machine learning on classical data has been receiving substantial attention \cite{tfq, pennylane, Liu, Chen, Chen2, qntk}. In particular,  PQCs have been emphasized because of their feasibility on  near-term devices. These approaches are also fraught with theoretically proven challenges, however. For quantum machine learning to be advantageous, a significant geometric difference between the classical and quantum data embedding is needed, and the quantum embedding must map to the correct answers with lower complexity than that of the classical embedding \cite{Huang}. Further, supervised quantum machine learning can be viewed as a kernel method, where the variational processing circuit is treated as a mere extension to the measurement, and the most important aspect is the embedding circuit once again \cite{schuld}. In light of this, Lloyd  et al.~\cite{lloyd} proposed a hybrid algorithm that learns the quantum embeddings that maximally separate different classes.

In this paper we propose a quantum few-shot embedding learning paradigm, which learns a mapping from classical data to quantum embeddings that are useful for training downstream quantum machine learning tasks. Crucially, we identify the \textit{circuit bypass problem} in hybrid neural networks, where learned classical parameters do not have large geometric differences from quantum embeddings. We observe that the few-shot learned embeddings generalize to unseen classes with high accuracy and suffer less from the circuit bypass problem compared with other approaches.

To enable this study, we develop a quantum machine learning library, QTensorAI, for large-scale machine learning studies. The plethora of numerical simulations and experiments that discuss potential quantum advantage investigate small system sizes because of simulation difficulties. This focus is especially problematic because  some learning algorithms with few qubits are known to perform better, but those with more than 10 to 15 qubits are significantly worse because the embedding schemes fail to realize large geometric differences for an exponentially large Hilbert space \cite{Huang, tfq}. We seamlessly integrate highly parallel tensor network quantum circuit simulations with the standard deep learning framework and enable simulations of circuits with large numbers of qubits. This capability allows us to verify our numerical results with confidence.

\section{Classical Few-Shot Learning}

Few-shot learning (FSL) is a machine learning paradigm that is designed to tackle a class of supervised learning problems  where the user has limited access to supervised data \cite{FSL}. This paradigm is potentially helpful in experimenting with human-like learning \cite{Lake}, learning from rare cases \cite{rare}, and reducing data-gathering effort and computational cost \cite{Fei}. The key distinction between FSL and other supervised learning methods is the use of  prior knowledge. Typical categorization of the FSL papers is based on how prior knowledge is used to help solve the problem \cite{FSL}:
\begin{itemize}
	\item Data approach: increases the number of samples (data augmentation like rotation, cropping, etc.)
	
    \item Model approach: constrains hypothesis space (multitask learning, embedding learning, etc.)
	
	\item Algorithm approach: alters search strategy in hypothesis space (refining existing parameters, refining metalearned parameters, learning the optimizer)
\end{itemize}

In this work we investigate prototypical networks,  a model approach that uses  embedding learning. This makes prototypical networks (and other embedding learning approaches) particularly relevant to our quantum embedding learning objective.

Consider a set of $N$ images
\begin{equation}
    \mathcal{D}=\{(\mathbf{x}_1,y_1),\dots,(\mathbf{x}_N,y_N)\}
\end{equation}
with $C$ classes \cite{proto}. We can choose $n<C$ classes randomly, sample $k$ images from each class $c$ to form a support set $S_c$, and sample a number of images to form a query set $Q_c$. The goal is to predict the label of images in $Q=\bigcup Q_c$ based on the labels in $S=\bigcup S_c$. We call this task $L$, which is an $n$-way $k$-shot classification task.

Since  many ways of sampling classes and images exist, $L\sim T$ follows a distribution of all possible tasks \cite{matching}. During validation, as long as the validation task is sampled from $T$, the algorithm should be able to correctly label images with only a few examples, even for classes that were never seen during training. We usually need a large number of image classes to produce many few-shot tasks. Overall, the algorithm must learn to \textit{extract features that are generally useful} for all tasks $L$, rather than those that are useful for only one specific task.

In prototypical networks \cite{proto} and matching networks \cite{matching}, embedding $\phi$ of individual images are extracted by using a neural network. For prototypical networks, a prototypical embedding is produced for each class $c$:
\begin{equation}
    \phi_c^p = \frac{1}{k}\sum_{i\in S_c} \phi_i .
\end{equation}

The query sample is classified in the class whose distance $d$ from the query embedding is minimized. The probability distribution of classifying a query sample $x^q$ as class $c$ is the softmax function over the distances between the query embedding and the prototypical embeddings of all classes: \cite{proto}
\begin{equation}
    p_\text{classical}(y^q=c|x^q)=\frac{\text{exp}(-d(\phi^p_c,\phi^q))}{\sum_{c'}\text{exp}(-d(\phi^p_{c'},\phi^q))},
\end{equation}
where $\phi^q$ is the query embedding. The loss $L$ is the negative log-probability of the true class $c$ \cite{proto}:
\begin{equation}
    L_\text{classical}=-\text{log}(p(y^q=c|x^q)).
\end{equation}

The Euclidean distance $d\in[0,\infty]$ and similar samples have smaller $d$. Softmax gives a target of $[0,1]$ and admits a probabilistic interpretation.

\section{Quantum Few-Shot Learning}

\begin{figure} [ht]
   \begin{center}
   \includegraphics[width=8cm]{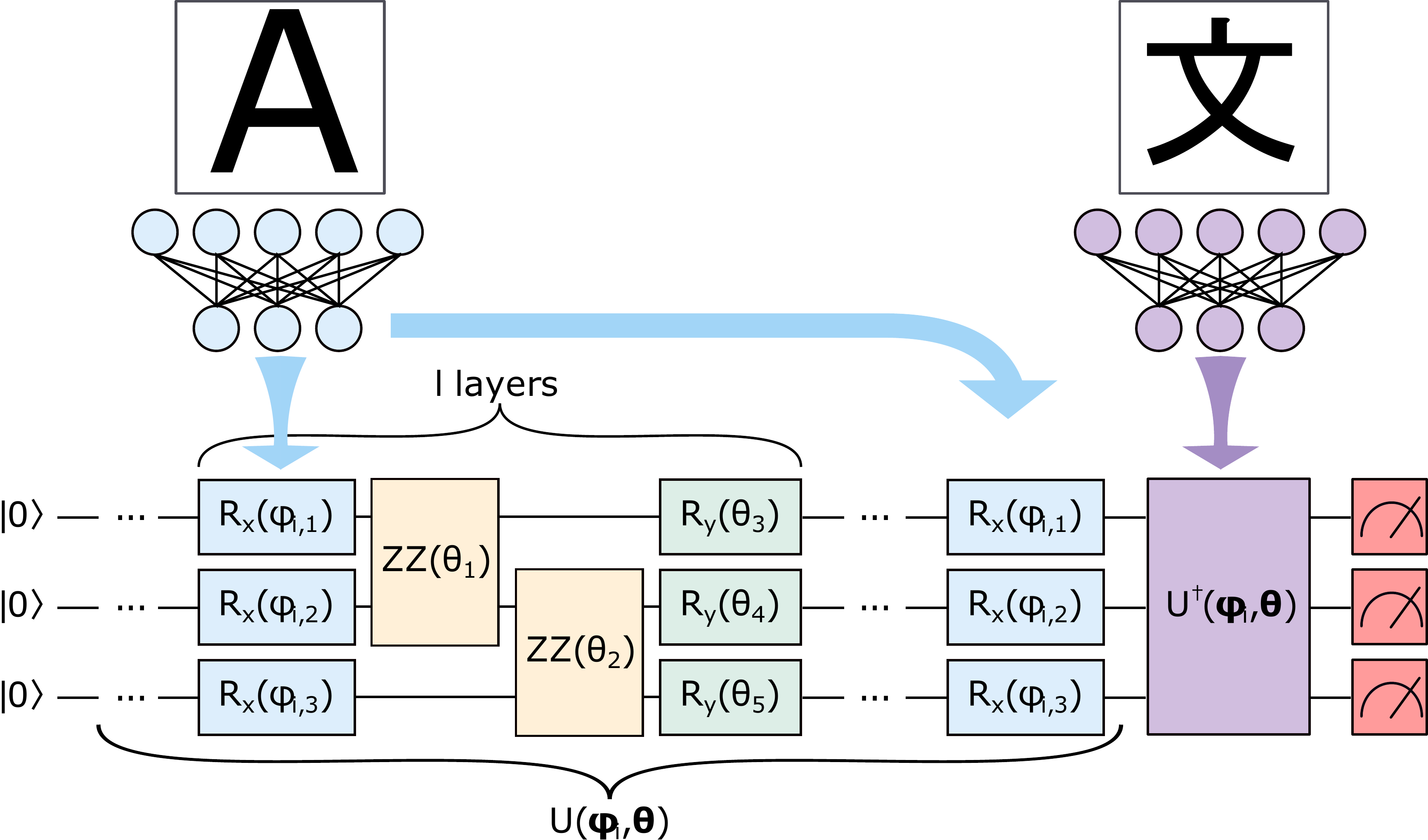}
   \end{center}
   \caption{Illustration of quantum FSL. Images of the two characters are processed with a classical neural network, and the outputs are used as the $\phi$ parameters of the metric learning quantum circuit. This circuit computes the fidelity between the quantum embeddings of the two characters using the inversion test, which measures the amplitude of the $|00\dots0\rangle$ state.}
   { \label{schematic} 
}
   \end{figure}

As we mentioned, learning quantum embeddings that are maximally separating is crucial, rather than relying on a fixed embedding scheme and training the processing circuit \cite{Huang, schuld, lloyd}. If we train the network on a specific task,  the learned embeddings are likely to  be task dependent and not very useful in other tasks. Classically, embedding learning can be achieved by using unsupervised learning techniques such as autoencoders \cite{Liou2008, Liou2014}, where a decoder that regenerates the original input from a bottlenecked encoding from an encoder can be used. Although quantum variants of autoencoders have been proposed \cite{romero, qvae}, they were not developed for learning quantum embeddings of classical data as far as we know. 

We propose FSL as a paradigm for quantum embedding learning because it samples hundreds or thousands of tasks and therefore the resulting embeddings are not task specific. Crucially, we observe that the learned embeddings do not form a low-dimensional manifold compared with classification or regression embeddings, which we will elaborate on in the \textit{Learned Pre-Quantum Embeddings} subsection in the \textit{Results} (\ref{embeddin_results}). Further, the learned embeddings can generalize to unseen classes. We emphasize that we do not claim quantum enhancement on FSL, although it is worth exploring in its own right. Instead, we focus on the few-shot enhancement of quantum embeddings and their limitations at this stage.

\subsection{Quantum Embedding}

To learn quantum embeddings of classical inputs, we replace the distance estimation part of the algorithm with a parameterized quantum circuit, as shown in Fig. \ref{schematic}. For each input sample $\mathbf{x}_{i}$, the classical neural network $\mathbf{f}_{cl}$, where all layers have ReLU activation except the last layer, processes the input up to the distance evaluation part, giving classical values
\begin{equation}
\boldsymbol\phi_{i}=\mathbf{f}_{cl}(\mathbf{x}_{i}).
\end{equation}
The classical values are used as rotation angles to the quantum circuit $U$,  translating a classical embedding of the sample into a quantum feature vector that is the state vector produced by the circuit. We call these learned classical rotation angles the \textit{prequantum embedding}.

Consider a quantum circuit that takes the prequantum embedding $\boldsymbol\phi_i$ and additional parameters $\boldsymbol\theta$ that are independent of the prequantum embeddings. The embedded quantum state is
\begin{equation}
    |\Psi(\mathbf{x}_i)\rangle=U(\boldsymbol\phi_i,\boldsymbol\theta)|0,\dots,0\rangle=U(\mathbf{f}_{cl}(\mathbf{x}_i),\boldsymbol\theta)|0,\dots,0\rangle.
\end{equation}

For the distance metric between the embedded samples, we use the fidelity between the corresponding quantum states:
\begin{equation}
d=F(|\Psi(\mathbf{x}_i)\rangle,|\Psi(\mathbf{x}_j)\rangle)=|\langle\Psi(\mathbf{x}_i)|\Psi(\mathbf{x}_j)\rangle|^2.
\end{equation}

In the prototypical network, however, the distance between the support samples and the query samples is \textit{not} evaluated pairwise. The network instead produces a prototypical vector of all support samples in a class, which is the average vector of all the vectors. Classically, the prototypical vector is trivial to obtain by taking the mean of all classical values. Since our embeddings are quantum state vectors, however, constructing a prototypical vector in the quantum embedding space is not easy. Naively taking the mean of the classical prequantum embedding would not result in a linear sum of the quantum feature vectors. However, since all we need is the inner product between the query vector and the class $c$ prototypical vector and that the inner product is linear, we can obtain the desired inner product by summing the individual inner products between all support vectors and the query vector:
\begin{equation}
    \langle\Psi^p_c|\Psi(\mathbf{x}^q)\rangle=\frac{1}{k}\sum_{i\in S_c}\langle\Psi(\mathbf{x}^s_i)|\Psi(\mathbf{x}^q)\rangle,
\end{equation}
where $|\Psi^p\rangle$ is the prototypical vector of the class, $\mathbf{x}^s_i$ is the $i$th support sample, and $\mathbf{x}^q$ is the query sample. If we want to obtain the final inner product, the full complex amplitude of the individual inner products must be obtained rather than just the absolute value. We can do so  by using a generalized Hadamard test, which is discussed in the \textit{Methods} section (\ref{Hadamard_test}). Note that if we use the matching network \cite{matching} paradigm of metalearning, we can use the absolute value of the pairwise inner products directly without having to worry about the complex amplitude, since it does not construct prototypical vectors.

\subsection{Loss Function}

For our quantum model, the inner product admits a natural probabilistic interpretation without the need for  softmax. The normalized classification probability and the loss are
\begin{align}
    p_\text{quantum}(y^q=c|x^q)&=\frac{|\langle\Psi^p_c|\Psi(\mathbf{x}^q)\rangle|^2}{\sum_{c'}|\langle\Psi^p_{c'}|\Psi(\mathbf{x}^q)\rangle|^2},\\ L_\text{quantum} &= -p_\text{quantum},
\end{align}
In fact, numerical simulations show that if we use the softmax probability and the negative log-probability loss, the performance is slightly worse.

\subsection{Quantum Circuit Ansatz}

We use the quantum circuit ansatz proposed by Lloyd  et al.~\cite{lloyd} to encode the prequantum embedding into quantum feature vectors. In their ansatz, they initialize all qubits to the $|0\rangle$ state, repeat an encoding layer $l$ times, and then add a layer of $R_x$ gates parameterized on the prequantum embedding. For the single encoding layer, they perform a single layer of $R_x$ gates whose angles are the prequantum embedding, followed by a chain of individually parameterized pairwise $ZZ$ gates (for our simulation, we  include only nearest-neighbor $ZZ$ interactions), and then a layer of individually parameterized $R_y$ gates. The parameters of the $ZZ$ and $R_y$ gates are independent of the prequantum embedding and stay the same across different inputs, but they are updated during training: they are parameters unique to the quantum circuit. We denote these parameters as $\boldsymbol\theta$.

We will call the ansatz proposed by Lloyd et al.~\cite{lloyd} the \textit{metric learning ansatz}. We also study the effects of initialization of the $\boldsymbol\theta$ parameters that determine how much entanglement the circuit has. We initialize $\boldsymbol\theta$ uniformly within specified ranges. When the initialization range is $0$, the circuit is equivalent to one made of single-qubit rotation gates only. We can increase this range up to $2\pi$ to turn up the circuit entanglement.

In the unentangled case, the resulting $n$-qubit state is
\begin{equation}
    |\Psi_i\rangle=\bigotimes_{\mu}^n (\text{cos}\phi_{i,\mu}|0\rangle-i\text{sin}\phi_{i,\mu}|1\rangle),
\end{equation}
up to  a multiplicative factor in the angles due to the sequential application of $R_y$ rotations in multiple layers. The inner product between two states corresponding to two samples is
\begin{equation}
    \langle\Psi_{i}|\Psi_j\rangle
    =\prod_{\mu}^n\text{cos}(\phi_{i,\mu}-\phi_{j,\mu}).
\end{equation}
This is a sensible result since we expect large inner products for inputs with similar angles. Moreover, we can encode $2^n$ mutually orthogonal states easily. Of course, since we are not using entanglement, this is just a quantum-inspired, classically tractable representation. The weakly entangled ansatz seeks to bring this to a more classically intractable representation.

Comparatively, the classical cosine distance can  encode only $n$ orthogonal vectors. For the Euclidean distance, we can encode infinitely many ``approximately orthogonal" vectors since we can simply spread them out in a  large (but only $n$-dimensional) space, and the negative exponential in the softmax function will make their confusion probability small.

\section{Hybrid Learning Results}

\subsection{Few-Shot Learning Benchmark}

For classical networks we examine the performance of both the Euclidean distance and the cosine distance metric. The Euclidean distance is known to outperform the cosine distance \cite{proto}; and since the quantum inner product is more similar to the cosine distance, we believe the cosine distance is a fairer comparison.

\begin{figure} [ht]
   \begin{center}
   \includegraphics[width=8cm]{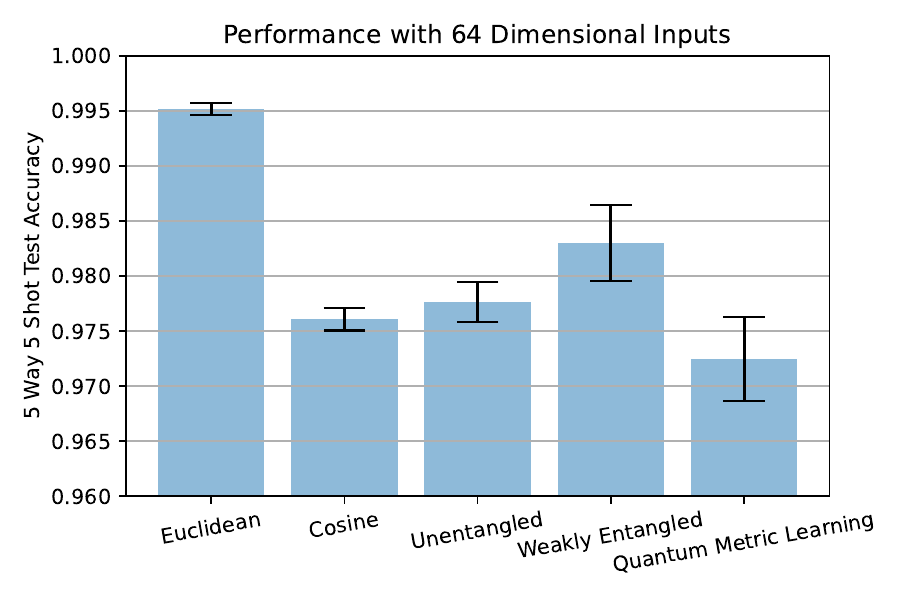}
   \end{center}
   \caption{Performance of models for the Omniglot dataset benchmark. All models are trained for 10 epochs. The highest testing accuracy during training is taken for each trial and averaged over 10 trials.} 
   { \label{omniglot} 
}
   \end{figure}

The dataset we use is the Omniglot \cite{omniglot} dataset. Omniglot consists of 1,623 handwritten characters from 50 alphabets with 20 examples each. We train on 60-way 5-shot tasks and test on 5-way 5-shot tasks, following the default settings of Snell  et al.~\cite{proto}. We evaluate the performance of models with prequantum embeddings of size 64 corresponding to 64 qubits. The performance comparison is shown in Fig. \ref{omniglot} Our best quantum model achieves a classification testing accuracy of over $98\%$, which is  significant. We stress that the testing accuracies are calculated on types of characters that were \textit{never} seen during training, indicating the great generalization capabilities of the quantum embedding scheme.

\begin{figure} [ht]
   \begin{center}
   \includegraphics[width=8cm]{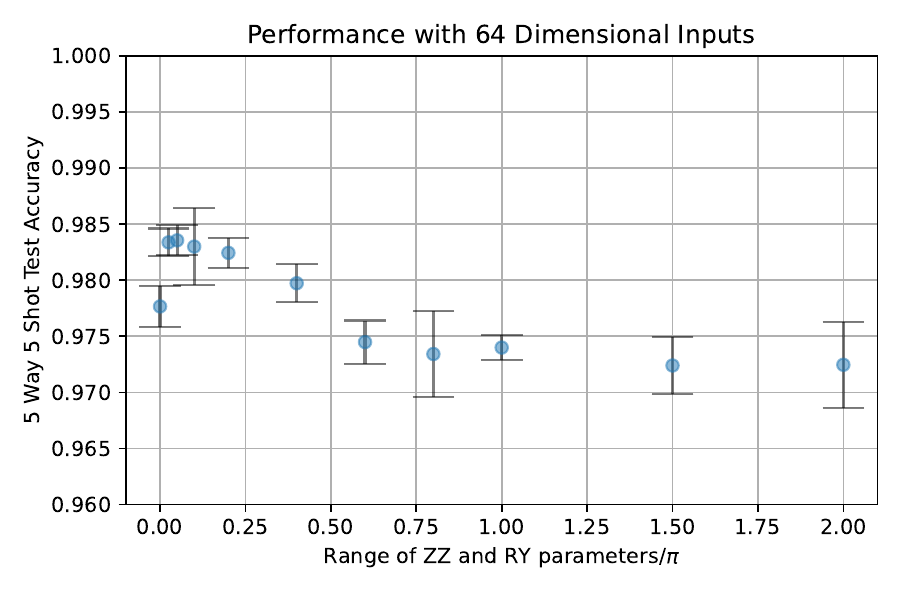}
   \end{center}
   \caption{Performance of models vs. the amount of entanglement at initialization. The $R_y$ and $ZZ$ gates are parameterized by $\mathbf{\theta}$, which are uniformly sampled from a range. The larger the range, the more entangled the circuit. Results are averaged over 3 trials except those already shown in Fig. \ref{omniglot}}
   { \label{entanglement} 
}
   \end{figure}

The quantum learning results are slightly worse than the classical counterpart that uses the Euclidean distance. This is not too surprising because the Euclidean distance allows the model to pack many clusters into even a single dimension, as long as they are separated. On the other hand, inner-product-type metrics can  distinguish between vectors only by angular separations. However, the quantum variant does outperform the classical cosine distance counterpart. In fact, we can see from Fig. \ref{entanglement} that as we turn on entanglement slightly, the model performance actually improves before it degrades again. Future work can potentially improve upon this with more intelligent ansatz design, initialization, and training procedures to better utilize the Hilbert space.

\subsection{Learned Prequantum Embeddings} \label{embeddin_results}

As we see from the results of quantum FSL, certain issues  prevent the efficient exploitation of the large Hilbert space of the quantum encoder. We seek to improve our understanding of the potential causes by examining the learned prequantum embeddings.

\subsubsection{Low-Dimensional Tasks}

\begin{figure} [ht]
   \begin{center}
   \includegraphics[width=6cm]{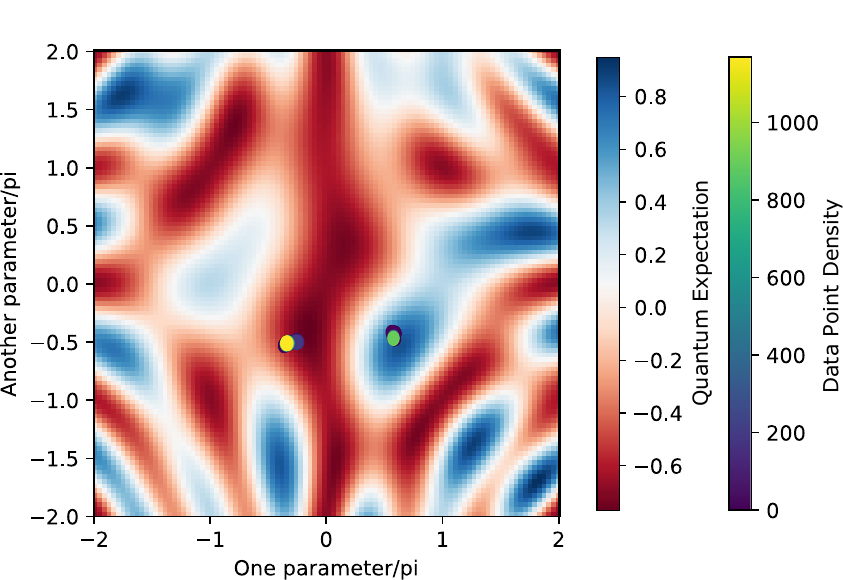}
   \end{center}
   \caption{Embedding data points on the quantum feature space for classification. The blue-red colored images spanning the whole domain are the feature spaces of the learned quantum circuit. The maps show the measurement expectation of the quantum circuit with different prequantum encoding inputs. The blue-yellow density scatter plot shows the density of prequantum embedding given all possible inputs.} 
   { \label{classification} 
}
   \end{figure}

\begin{figure} [ht]
   \begin{center}
   \includegraphics[width=6cm]{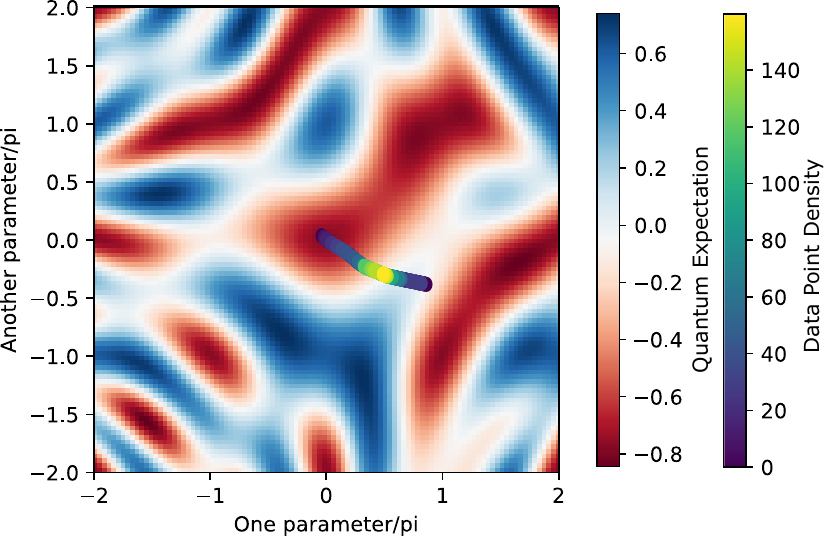}
   \end{center}
   \caption{Embedding data points in the quantum feature space for regression of one variable.} 
   { \label{1d} 
}
   \end{figure}

\begin{figure} [ht]
   \begin{center}
   \includegraphics[width=6cm]{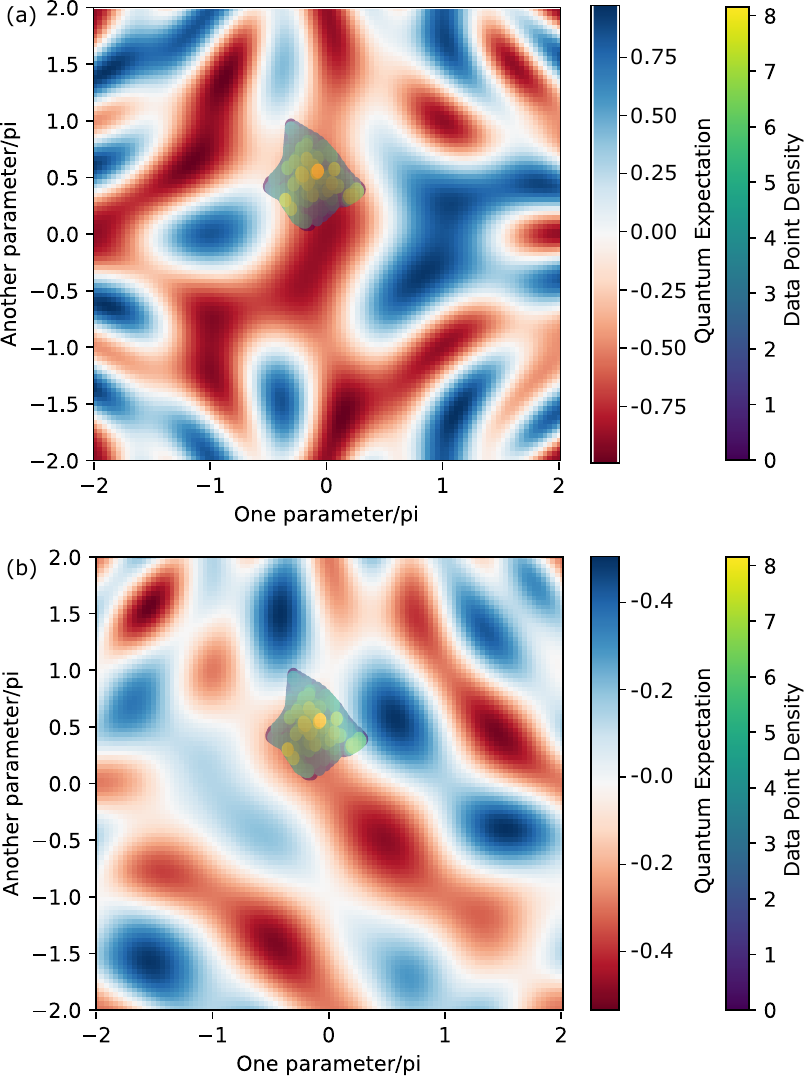}
   \end{center}
   \caption{Embedding data points in the quantum feature space for regression of two variables.} 
   { \label{2d} 
}
   \end{figure}

For simplicity, we first consider the prequantum embeddings learned during classification, single-variable regression, and two-variable regression tasks, not FSL. Details of implementation can be found in thee \textit{Methods} section (\ref{ld}). The results are shown in Fig. \ref{classification}, \ref{1d}, and \ref{2d}, respectively.

For the classification task, we find that the prequantum embeddings are concentrated in two spots. Each spot corresponds to a different label. This shows that the classical neural network learns to classify the samples and simply outputs the values of prequantum embedding that would give rise to the correct measurement value.

For the single-variable regression task, the prequantum embeddings form a line. Moreover, the line is over a region in the feature space such that the expected measurement outcome would span a range of values, which is needed for regression: we note that in this and other trials, the prequantum embeddings tend to cross white regions, going from blue to red. This shows that the classical network learns to output the prequantum embedding that would lead to the correct prediction.

For the two-variable regression task, we use two quantum circuits, each responsible for one component of the prediction values. The interpretation of the learned prequantum embedding is harder. We expect, however,  that it needs to occupy a two-dimensional region to be able to give distinct two-dimensional outputs.

Overall, we observe that if the target output is $n$  dimensional, the learned prequantum embedding will occupy an $n$-dimensional manifold. This is especially problematic if we hope to exploit quantum neural networks for potential advantage. Essentially, the classical neural network  cares only about getting the quantum neural network to give the correct output, and not the utilization of the quantum circuit, essentially bypassing the circuit and  treating it as a strange nonlinearity. We refer to this as the \textit{circuit bypass problem}. The low dimensionality of the prequantum embedding space is actually shown in Fig. 5 of Lloyd  et al.~\cite{lloyd} but was not viewed a problem at the time. In their work, the learned prequantum embeddings form 0-D dots for classification, which illustrates that our statement about the circuit bypass problem also appears in more complex architectures and problems.

More formally, let the output space be $\mathcal{Y}\in \mathbb{R}^n$, and assume that the trained prequantum embedding point clouds in the $d>n$-dimensional space $\mathcal{E}\in\mathbb{R}^d$ are sampled with the probability measure $\mathcal{P}$ supported on the unit ball \cite{Fefferman}. We hypothesize that for hybrid networks trained without special treatments,  a submanifold 
$\mathcal{M}\in\mathcal{G}(n,CV,\tau/C)$ exists such that $\mathcal{L(M,P)}<C\epsilon$, where $\mathcal{L(M,P)}=\int d(x,\mathcal{M})^{2}\mathbf{d}\mathcal{P}(x)$ measures the error of the manifold from the distribution, $d(x,\mathcal{M})=\text{inf}_{y\in\mathcal{M}}|x-y|$, $\mathcal{G}(n,CV,\tau/C)$ is the family of $n$-dimensional submanifolds of the unit ball in $\mathcal{E}$ with volume $\leq CV$ and reach $\geq\tau/C$, $V$ and $\tau$ are hyperparameters that constrains the complexity of the manifolds, and $C$ is a $d$-dependent constant. In other words, the prequantum embeddings approximately occupy an $n$-dimensional manifold.

Further, for the smallest $\mathcal{M}$ satisfying the condition, $f:\mathcal{M}\rightarrow\mathcal{Y(M)}$ is likely a diffeomorphism, where $f$ is a function from the manifold to the output space, with points from the manifold as inputs. This is in contrast with the condition for quantum advantage in learning with quantum-enhanced feature spaces, where many-to-one mappings are characteristic.

\subsubsection{Few-Shot Learning}
 
   \begin{figure} [ht]
   \begin{center}
   \includegraphics[width=6cm]{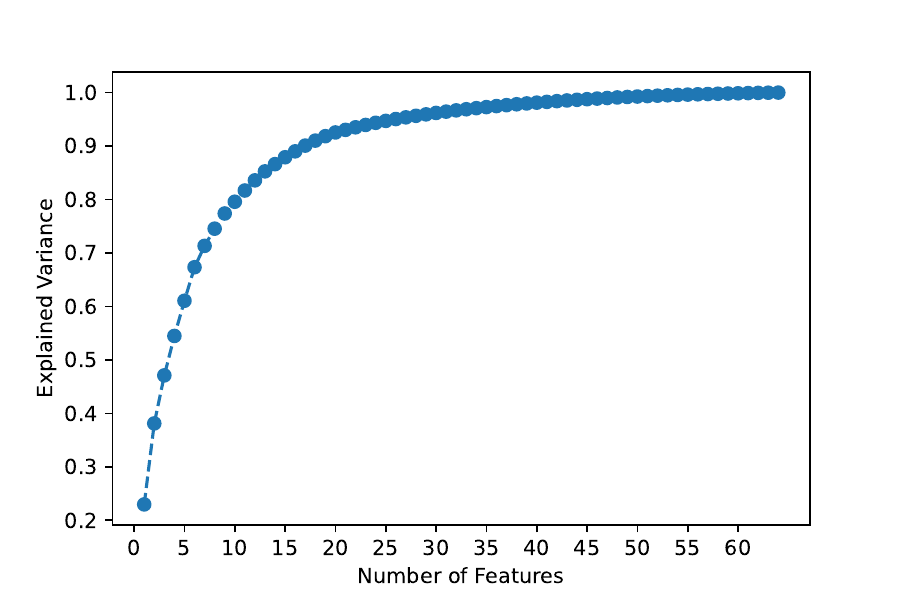}
   \end{center}
   \caption{Explained variance against the number of components for prequantum embeddings.} 
   { \label{pca} 
}
   \end{figure}
 
We now return to few-shot learning. In general, the need to distinguish hundreds of classes of images means that the learned embeddings should maximally occupy the embedding space as much as possible to reduce confusion and is therefore generally on a high-dimensional manifold (limited by the dimension of the embedding space). Indeed, we verify that the learned prequantum embeddings for the Omniglot dataset do nontrivially occupy multiple dimensions. We perform principal component analysis on the prequantum embeddings and show in Fig. \ref{pca} the explained variance against the number of components. We observe that around 15 dimensions are required to explain $90\%$ of the variance, indicating that the embeddings are not low dimensional.
 
\begin{figure} [ht]
   \begin{center}
   \includegraphics[width=8cm]{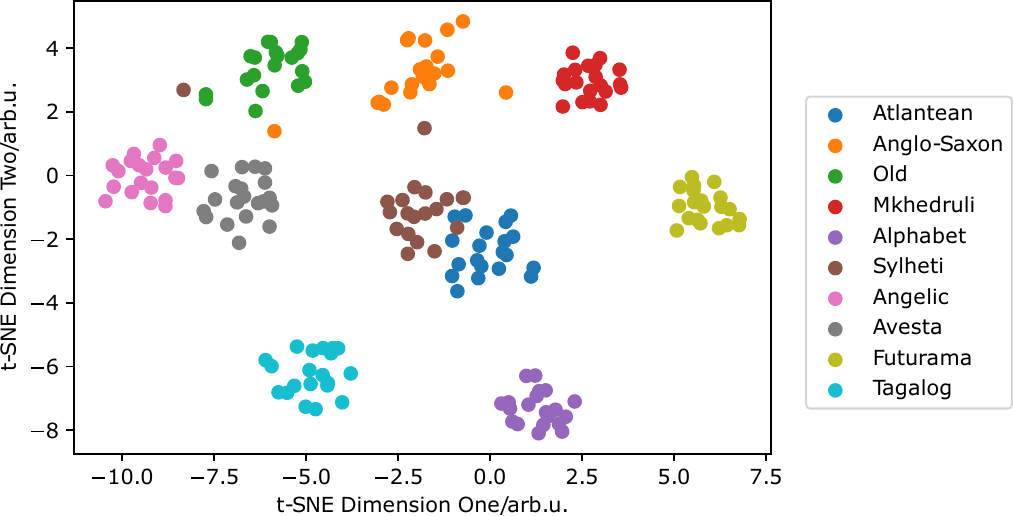}
   \end{center}
   \caption{Prequantum feature space embeddings for FSL.} 
   { \label{quantum} 
}
   \end{figure}

\begin{figure} [ht]
   \begin{center}
   \includegraphics[width=8cm]{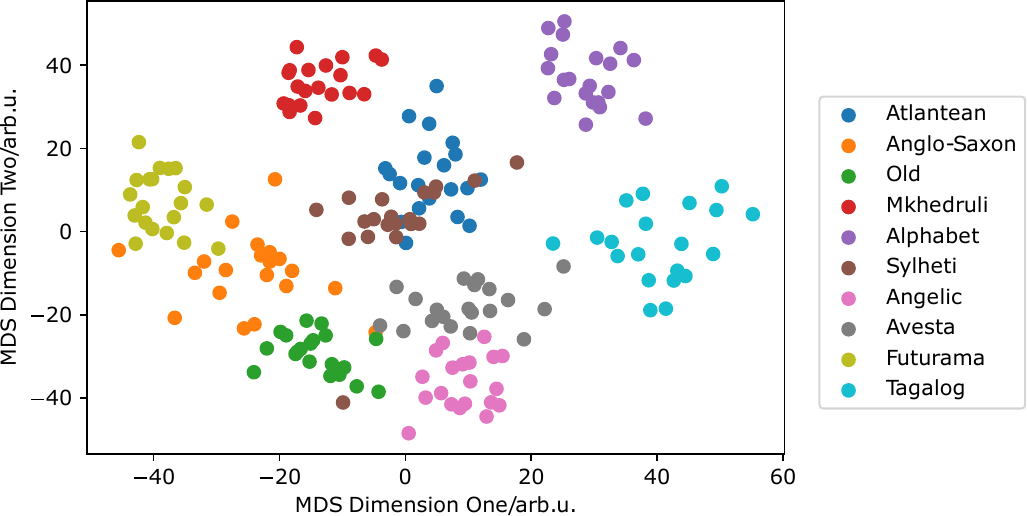}
   \end{center}
   \caption{Quantum feature space embeddings for FSL.} 
   { \label{pre_quantum} 
}
   \end{figure}

To examine the embedding in depth, we choose 10 classes of characters randomly and visualize in Fig. \ref{pre_quantum} their embeddings in 2D utilizing t-SNE \cite{tsne}. We observe that the prequantum embeddings form clusters according to their class assignment. To examine the quantum embeddings, instead of computing the Hilbert space state vectors directly, which is prohibitively difficult for 64 qubits, we project the quantum embeddings onto a 2D plane according to their pairwise distances using MDS \cite{mds} in Fig. \ref{quantum}. Specifically, we use the negative logarithm of the square of the inner product as pairwise dissimilarity. One can clearly see that the structure of the prequantum embedding space is almost preserved in the quantum embedding space. This situation cannot admit large geometric differences between the quantum embedding and the classical embedding, hence ruling out quantum advantage.

We believe this problem is similar  to the low-dimensional manifold issue. Since the quantum circuit preserves the geometric relationship between data points  well, the function $f:\mathcal{M}\rightarrow \mathcal{H}(\mathcal{M})$ is likely a diffeomorphism, where $\mathcal{M}$ is the smallest approximate manifold for the prequantum embeddings and $\mathcal{H}$ is the quantum Hilbert space. Therefore, we also identify this issue under the circuit bypass problem.
 
\section{Methods}

\subsection{High-Performance Simulation}

In deep learning, training the neural network requires thousands or millions of samples, and parallel computation is often necessary. Typically, a batch of training samples is passed to the neural network and processed in parallel to speed up the training. Training quantum neural networks is even harder because of the exponential cost in simulating quantum circuits as the number of qubits grows. We employ two approaches to speed up quantum neural networks: the tensor network approach and batch parallelism for quantum simulation. The result is that we incur a linear cost in the number of qubits for shallow circuits and are able to efficiently utilize all GPU resources available.

\subsubsection{Tensor Network Simulator}

Tensor network simulators treat quantum circuits as a graph \cite{Boixo}. The most straightforward way to use a tensor network simulator is to evaluate the transition amplitude of the circuit unitary between two computational basis states. Evaluation of such amplitudes corresponds to contracting the tensor network into a rank zero tensor. In the limit of large $n$ and small $l$, the tensor network would have a small transverse size in the layer direction and can be contracted in the longitudinal qubit direction to minimize the largest rank of the tensor that needs to be stored. Since the memory cost is exponential to the rank of the tensor, finding the contraction order that minimizes the rank is crucial \cite{Schutski} and does not generally boil down to heuristics such as contracting along the qubit direction.

Our model evaluates the fidelity between two quantum states prepared by the encoding circuit, namely,
\begin{align}
    d&=F=|\langle|\Psi(\mathbf{x}_i)|\Psi(\mathbf{x}_j)\rangle|^2\\
    &=|\langle0,\dots,0|U^{\dagger}(\boldsymbol\phi_i,\boldsymbol\theta)U(\boldsymbol\phi_j,\boldsymbol\theta)|0,\dots,0\rangle|^2,
\end{align}
which is computable from exactly the transition amplitude between the $|0,\dots,0\rangle$ states when joining the encoding circuit of a sample $j$ with the inverse encoding circuit of a sample $i$. For this work we use QTensor \cite{Lykov2020, Lykov2021}, a tensor network simulator platform for  studies of the Quantum Approximate Optimization Algorithm. (QAOA) in high-performance computing environments. It is based on the QTree simulator.

\subsubsection{Batch Parallelism and GPU Optimization for AI}

QTensor was designed to evaluate a single large circuit at a time. In deep learning tasks we are often limited to smaller circuits but need to evaluate them with different parameterization across a whole batch or other parallel dimensions (i.e., those due to convolution). QTensor is already equipped with backpropagation because of the need to update QAOA parameters, and it has a PyTorch backend to support that. Therefore, it is in principle friendly to parallelization, and we developed the library QTensorAI to achieve parallelism and convenient interfacing with PyTorch models by wrapping components of the quantum neural network as \textit{nn.Module} objects. Because the same circuit for a given module will be executed thousands or even millions of times, it is advantageous to optimize the circuit contraction order to aggressively minimize the maximum tensor rank upon model initialization and then reuse this contraction order for future executions.

Moreover, with the use of PyTorch CUDA Graphs \cite{cudagraph, pytorchcudagraph}, we further speed up the simulations. For circuits that need many contraction steps, the CPU needs to launch GPU kernels hundreds or thousands of times, incurring significant overhead. Tracing CUDA Graphs allows static graphs to be captured and replayed, reducing the CPU call to a single graph launch operation.

\subsection{Details of Low-Dimensional Tasks} \label{ld}

All classical neural networks for low-dimensional tasks have $8$ fully connected layers with ReLU activation except the layer before the quantum circuit. For classification or regression tasks, we need not limit ourselves to a kernel approach and use the metric learning ansatz, which requires pairwise evaluation. Instead, all the circuits used here have two qubits and follow the ansatz used by Abbas  et al.~\cite{Abbas}, where outputs are measurement expectations. Specifically, two input values from classical neural networks are fed to the encoding circuit, followed by three variational layers whose parameters are separately optimized.

We use two-dimensional inputs for all the tasks. For the classification task, if the first component is larger than the second, we assign the label $1$. Otherwise, it is assigned $-1$. For the single-variable regression task, the target value is proportional to the difference between the two components. For the two-variable regression task, one target value is proportional to the difference, and the other is proportional to the sum between the two components.

\subsection{Hadamard Test} \label{Hadamard_test}

\begin{figure} [ht]
   \begin{center}
   \includegraphics[width=4cm]{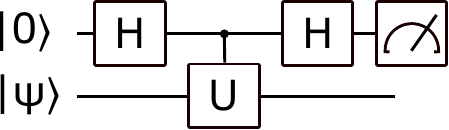}
   \end{center}
   \caption{Hadamard test circuit.} 
   { \label{Hadamard_test_circuit} 
}
   \end{figure}

To obtain the complex value of the inner product of the quantum embeddings on a real device, we can use the Hadamard test \cite{aharonov2009polynomial}. The test estimates Re$\langle\psi|\mathcal{U}|\psi\rangle$ (or Im$\langle\psi|\mathcal{U}|\psi\rangle$ with modification) with the circuit shown in Fig. \ref{Hadamard_test_circuit}. The estimate is given by the probability difference between measuring the ancilla in $|0\rangle$ and in $|1\rangle$.

In our case, \begin{eqnarray}
    |\psi\rangle=|0,\dots,0\rangle\\ \mathcal{U}=U^{\dagger}(\boldsymbol\phi_i,\boldsymbol\theta)U(\boldsymbol\phi_j,\boldsymbol\theta).
\end{eqnarray}

Further, to implement controlled $U$ for the metric learning ansatz, we  need only single-qubit controlled gates: we have controlled $R_x$ gates that are $\phi$-dependent, and the $\boldsymbol\theta$ dependent $ZZ$ and $R_y$ gates will cancel out if the ancilla is in $|0\rangle$ and the $R_x$ gates are not executed.

\section{Conclusion and Outlook}

We investigated the quantum few-shot learning paradigm as a method of learning general quantum embeddings. We find that our weakly entangled quantum embeddings exhibit superior performance over classical cosine distance. Although this result does not offer quantum enhancement to few-shot learning since the Euclidean distance yields better results, the quantum embeddings obtained with few-shot learning have the most desirable properties compared with alternatives. First, the learned embeddings generalize  well to unseen classes. Second, quantum few-shot embeddings do not occupy low-dimensional manifolds compared with classification-based embedding learning methods.

We identified the \textit{circuit bypass problem} that can plague hybrid quantum embedding learning approaches and other hybrid architectures, making them unable to realize high geometric differences and, consequently, quantum advantage. However, methods to regularize the prequantum embedding space by penalizing the formation of low-dimensional manifolds or maximizing the geometric distance could resolve this issue. Recently, a regularization that encourages low-dimensional manifolds has been developed that calculates the manifold dimension and adds that to the loss function \cite{LDMNet}. In our setting we can subtract the manifold dimension instead to force the embeddings to occupy high-dimensional manifolds. Further, the geometric difference between the classical prequantum and the quantum embedding can be explicitly computed and added to the loss function, encouraging nontrivial quantum encodings. These directions hold the potential to mitigate the circuit bypass problem and warrant future investigation.

\section{Acknowledgment}

This research used resources of the Argonne Leadership Computing Facility, which is a DOE Office of Science User Facility supported under Contract DE-AC02-06CH11357. This research also used the Princeton Research Computing resources at Princeton University which is consortium of groups led by the Princeton Institute for Computational Science and Engineering (PICSciE) and Office of Information Technology's Research Computing. This material is based upon work supported by the U.S. Department of Energy, Office of Science, under contract number DE-AC02-06CH11357.  We thank William Tang for the helpful discussions.

\section{Code availability}
The QTensorAI library is open source available in the GitHub repository \cite{qtensor_ai}. The quantum few-shot learning repository is available in the GitHub repository \cite{QFSL}.

\bibliographystyle{IEEEtran}
\bibliography{main.bib}

\begin{thebibliography}{10}
\providecommand{\url}[1]{#1}
\csname url@samestyle\endcsname
\providecommand{\newblock}{\relax}
\providecommand{\bibinfo}[2]{#2}
\providecommand{\BIBentrySTDinterwordspacing}{\spaceskip=0pt\relax}
\providecommand{\BIBentryALTinterwordstretchfactor}{4}
\providecommand{\BIBentryALTinterwordspacing}{\spaceskip=\fontdimen2\font plus
\BIBentryALTinterwordstretchfactor\fontdimen3\font minus
  \fontdimen4\font\relax}
\providecommand{\BIBforeignlanguage}[2]{{%
\expandafter\ifx\csname l@#1\endcsname\relax
\typeout{** WARNING: IEEEtran.bst: No hyphenation pattern has been}%
\typeout{** loaded for the language `#1'. Using the pattern for}%
\typeout{** the default language instead.}%
\else
\language=\csname l@#1\endcsname
\fi
#2}}
\providecommand{\BIBdecl}{\relax}
\BIBdecl

\bibitem{Du}
Y.~Du, M.~Hsieh, T.~Liu, and D.~Tao, ``Expressive power of parametrized quantum
  circuits,'' \emph{Phys. Rev. Research}, vol.~2, p. 033125, 2020.

\bibitem{Abbas}
A.~Abbas, D.~Sutter, and C.~Zoufal~{\sl et al}, ``The power of quantum neural
  networks,'' \emph{Nat. Comput. Sci.}, vol.~1, pp. 403--409, 2021.

\bibitem{Huang}
H.~Huang, M.~Broughton, and M.~Mohseni~{\sl et al.}, ``Power of data in quantum
  machine learning,'' \emph{Nat. Commun.}, vol.~12, p. 2631, 2021.

\bibitem{Cong}
I.~Cong, S.~Choi, and M.~Lukin, ``Quantum convolutional neural networks,''
  \emph{Nat. Phys.}, vol.~15, pp. 1273--1278, 2019.

\bibitem{qsvm}
P.~Rebentrost, M.~Mohseni, and S.~Lloyd, ``Quantum support vector machine for
  big data classification,'' \emph{Phys. Rev. Lett.}, vol. 113, no.~13, p.
  130503, 2014.

\bibitem{hhl}
A.~W. Harrow, A.~Hassidim, and S.~Lloyd, ``Quantum algorithm for linear systems
  of equations,'' \emph{Phys. Rev. Lett.}, vol. 103, p. 150502, Oct 2009.

\bibitem{schuld}
M.~Schuld, ``Supervised quantum machine learning models are kernel methods,''
  \emph{arXiv preprint arXiv:2101.11020}, unpublilshed.

\bibitem{Havlicek}
V.~Havlíček, A.~Córcoles, and K.~Temme~et al., ``Supervised learning with
  quantum-enhanced feature spaces,'' \emph{Nature}, vol. 567, pp. 209--212,
  2019.

\bibitem{tfq}
M.~Broughton, G.~Verdon, T.~McCourt, A.~J. Martinez, J.~H. Yoo, S.~V. Isakov,
  P.~Massey, R.~Halavati, M.~Y. Niu, A.~Zlokapa \emph{et~al.}, ``Tensorflow
  quantum: A software framework for quantum machine learning,'' \emph{arXiv
  preprint arXiv:2003.02989}, 2020.

\bibitem{pennylane}
V.~Bergholm, J.~Izaac, M.~Schuld, C.~Gogolin, M.~S. Alam, S.~Ahmed, J.~M.
  Arrazola, C.~Blank, A.~Delgado, S.~Jahangiri \emph{et~al.}, ``Pennylane:
  Automatic differentiation of hybrid quantum-classical computations,''
  \emph{arXiv preprint arXiv:1811.04968}, 2020.

\bibitem{Liu}
L.~Junhua, L.~Kwan~Hui, W.~Kristin~L., H.~Wei, G.~Chu, and H.~He-Liang,
  ``Hybrid quantum-classical convolutional neural networks,'' \emph{Sci. China:
  Phys. Mech. Astron.}, vol.~64, p. 290311, 2021.

\bibitem{Chen}
S.~Y.-C. Chen, C.-H.~H. Yang, J.~Qi, P.-Y. Chen, X.~Ma, and H.-S. Goan,
  ``Variational quantum circuits for deep reinforcement learning,'' \emph{IEEE
  Access}, vol.~8, pp. 141\,007--141\,024, 2020.

\bibitem{Chen2}
\BIBentryALTinterwordspacing
S.~Y.-C. Chen, T.-C. Wei, C.~Zhang, H.~Yu, and S.~Yoo, ``Quantum convolutional
  neural networks for high energy physics data analysis,'' \emph{Phys. Rev.
  Research}, vol.~4, p. 013231, March 2022. [Online]. Available:
  \url{https://link.aps.org/doi/10.1103/PhysRevResearch.4.013231}
\BIBentrySTDinterwordspacing

\bibitem{qntk}
J.~Liu, F.~Tacchino, J.~R. Glick, L.~Jiang, and A.~Mezzacapo, ``Representation
  learning via quantum neural tangent kernels,'' \emph{arXiv preprint
  arXiv:2111.04225}, 2021.

\bibitem{lloyd}
S.~Lloyd, M.~Schuld, A.~Ijaz, J.~Izaac, and N.~Killoran, ``Quantum embeddings
  for machine learning,'' \emph{arXiv preprint arXiv:2001.03622}, 2021.

\bibitem{FSL}
Y.~Wang, Q.~Yao, J.~T. Kwok, and L.~M. Ni, ``Generalizing from a few examples:
  A survey on few-shot learning,'' \emph{ACM Comput. Surv.}, vol.~53, no.~3,
  jun 2020.

\bibitem{Lake}
B.~M. Lake, R.~Salakhutdinov, and J.~B. Tenenbaum, ``Human-level concept
  learning through probabilistic program induction,'' \emph{Science}, vol. 350,
  no. 6266, pp. 1332--1338, 2015.

\bibitem{rare}
H.~Altae-Tran, B.~Ramsundar, A.~S. Pappu, and V.~Pande, ``Low data drug
  discovery with one-shot learning,'' \emph{ACS Central Science}, vol.~3,
  no.~4, pp. 283--293, 2017.

\bibitem{Fei}
L.~Fei-Fei, R.~Fergus, and P.~Perona, ``One-shot learning of object
  categories,'' \emph{IEEE Trans. Pattern Anal. Mach. Intell.}, vol.~28, no.~4,
  pp. 594--–611, apr 2006.

\bibitem{proto}
J.~Snell, K.~Swersky, and R.~Zemel, ``Prototypical networks for few-shot
  learning,'' in \emph{Advances in Neural Information Processing Systems},
  vol.~30.\hskip 1em plus 0.5em minus 0.4em\relax Curran Associates, Inc.,
  2017.

\bibitem{matching}
O.~Vinyals, C.~Blundell, T.~Lillicrap, k.~kavukcuoglu, and D.~Wierstra,
  ``Matching networks for one shot learning,'' in \emph{Advances in Neural
  Information Processing Systems}, vol.~29.\hskip 1em plus 0.5em minus
  0.4em\relax Curran Associates, Inc., 2016.

\bibitem{Liou2008}
C.-Y. Liou, J.-C. Huang, and W.-C. Yang, ``Modeling word perception using the
  {Elman} network,'' \emph{Neurocomput.}, vol.~71, no. 16–18, p. 3150–3157,
  oct 2008.

\bibitem{Liou2014}
C.-Y. Liou, W.~C. Cheng, J.-W. Liou, and D.-R. Liou, ``Autoencoder for words,''
  \emph{Neurocomput.}, vol. 139, pp. 84--96, 2014.

\bibitem{romero}
J.~Romero, J.~P. Olson, and A.~Aspuru-Guzik, ``Quantum autoencoders for
  efficient compression of quantum data,'' \emph{Quantum Sci. Technol.},
  vol.~2, no.~4, p. 045001, 2017.

\bibitem{qvae}
A.~Khoshaman, W.~Vinci, B.~Denis, E.~Andriyash, H.~Sadeghi, and M.~H. Amin,
  ``Quantum variational autoencoder,'' \emph{Quantum Sci. Technol.}, vol.~4,
  no.~1, p. 014001, 2018.

\bibitem{omniglot}
B.~M. Lake, R.~Salakhutdinov, J.~Gross, and J.~B. Tenenbaum, ``One shot
  learning of simple visual concepts,'' \emph{Cogn. Sci.}, vol.~33, 2011.

\bibitem{Fefferman}
C.~Fefferman, S.~Mitter, and H.~Narayanan, ``\BIBforeignlanguage{English
  (US)}{Testing the manifold hypothesis},'' \emph{\BIBforeignlanguage{English
  (US)}{J. Amer. Math. Soc.}}, vol.~29, no.~4, pp. 983--1049, 2016.

\bibitem{tsne}
L.~van~der Maaten and G.~Hinton, ``Visualizing high-dimensional data using
  {t-SNE},'' \emph{J. Mach. Learn. Res}, 2008.

\bibitem{mds}
I.~Borg and P.~J.~F. Groenen, \emph{Modern Multidimensional Scaling Theory and
  Applications}.\hskip 1em plus 0.5em minus 0.4em\relax New York: Springer,
  2005.

\bibitem{Boixo}
S.~Boixo, S.~V. Isakov, V.~N. Smelyanskiy, and H.~Neven, ``Simulation of
  low-depth quantum circuits as complex undirected graphical models,''
  \emph{arXiv preprint arXiv:1712.05384}, 2017.

\bibitem{Schutski}
R.~Schutski, D.~Lykov, and I.~Oseledets, ``Adaptive algorithm for quantum
  circuit simulation,'' \emph{Phys. Rev. A}, vol. 101, p. 042335, Apr 2020.

\bibitem{Lykov2020}
D.~Lykov, R.~Schutski, A.~Galda, V.~Vinokur, and Y.~Alexeev, ``Tensor network
  quantum simulator with step-dependent parallelization,'' \emph{arXiv preprint
  arXiv:2012.02430}, 2020.

\bibitem{Lykov2021}
D.~Lykov, A.~Chen, H.~Chen, K.~Keipert, Z.~Zhang, T.~Gibbs, and Y.~Alexeev,
  ``Performance evaluation and acceleration of the {QTensor} quantum circuit
  simulator on gpus,'' in \emph{2021 IEEE/ACM Second International Workshop on
  Quantum Computing Software (QCS)}, 2021, pp. 27--34.

\bibitem{cudagraph}
S.~Narasimhan, ``{Global Computer Makers Deliver Breakthrough MLPERF results
  with Nvidia AI},''
  \url{https://blogs.nvidia.com/blog/2021/06/30/mlperf-ai-training-partners/},
  accessed: 2022-03-07.

\bibitem{pytorchcudagraph}
V.~Nguyen, M.~Carilli, S.~Burc~Eryilmaz, V.~Singh, M.~Lin, N.~Gimelshein,
  A.~Desmaison, and E.~Yang, ``{Accelerating PyTorch with CUDA Graphs},''
  \url{https://pytorch.org/blog/accelerating-pytorch-with-cuda-graphs/},
  accessed: 2022-03-07.

\bibitem{aharonov2009polynomial}
D.~Aharonov, V.~Jones, and Z.~Landau, ``A polynomial quantum algorithm for
  approximating the {Jones} polynomial,'' \emph{Algorithmica}, vol.~55, no.~3,
  pp. 395--421, 2009.

\bibitem{LDMNet}
W.~Zhu, Q.~Qiu, J.~Huang, A.~R. Calderbank, G.~Sapiro, and I.~Daubechies,
  ``Ldmnet: Low dimensional manifold regularized neural networks,'' \emph{2018
  IEEE/CVF Conference on Computer Vision and Pattern Recognition}, pp.
  2743--2751, 2018.

\bibitem{qtensor_ai}
\BIBentryALTinterwordspacing
Mar 2022. [Online]. Available: \url{https://github.com/sss441803/QTensorAI}
\BIBentrySTDinterwordspacing

\bibitem{QFSL}
\BIBentryALTinterwordspacing
Mar 2022. [Online]. Available: \url{https://github.com/sss441803/QFSL}
\BIBentrySTDinterwordspacing

\end{thebibliography}

\end{document}